# Status of the Compressed Baryonic Matter experiment at FAIR

Peter Senger[1,2] for the CBM collaboration

*1 Facility for Antiproton and Ion Research, Darmstadt, Germany*

*2 National Research Nuclear University MEPhI, Moscow, Russia*

[p.senger@gsi.de](mailto:p.senger@gsi.de)

**Abstract:**

The Compressed Baryonic Matter (CBM) experiment will investigate high-energy heavy-ion collisions at the international Facility for Antiproton and Ion Research (FAIR), which is under construction in Darmstadt, Germany. The CBM research program is focused on the exploration of QCD matter at neutron star core densities, such as study of the equation-of-state and the search for phase transitions. Key experimental observables include (multi-) strange (anti-) particles, electron-positron pairs and dimuons, particle correlations and fluctuations, and hyper-nuclei. In order to measure these diagnostic probes multi-differentially with unprecedented precision, the CBM detector and data acquisition systems are designed to run at reaction rates up to 10 MHz. This requires the development of fast and radiation hard detectors and readout electronics for track reconstruction, electron and muon identification, time-of-flight determination, and event characterization. The data are read-out by ultra-fast, radiation-tolerant, and free-streaming front-end electronics, and then transferred via radiation-hard data aggregation units and high-speed optical connections to a high-performance computing center. A fast and highly parallelized software will perform online track reconstruction, particle identification, and event analysis. The components of the CBM experimental setup will be discussed, and results of physics performance studies will be presented.

## 1. The future Facility for Antiproton and Ion Research (FAIR)

The future international Facility for Antiproton and Ion Research (FAIR) in Darmstadt will open a new era of forefront research in nuclear physics, hadron physics, plasma physics, atomic physics, radiation biophysics, and material research. The FAIR start version will comprise the synchrotron SIS100, a collector ring, a storage ring, and a variety of experimental setups. SIS100 will provide high-intensity primary beams of protons up to 29 GeV, and nuclear beams with kinetic energies 15A GeV. Beams of rare isotopes will be selected from nuclear reactions by the Superconducting Fragment Separator, and guided for further investigated to the experimental facilities built and operated by the NUSTAR collaboration. The High-Energy Storage Ring (HESR) will accelerate and cool intense secondary beams of antiprotons, and focus them on the target of the PANDA detector, where hadron physics experiments will be performed. A dedicated cave will host the various detector setups for atomic, plasma, biophysics and material research. The Compressed Baryonic Matter (CBM) detector is designed for the investigation of the properties of dense nuclear matter, which will be created in energetic nucleus-nucleus collisions. Civil construction is well under way, and the manufacturing and test of the accelerator components is progressing. Installation and commissioning of the experiments is planned for 2022-2024, and the first beams from SIS100 are expected for 2025. The layout of FAIR is sketched in figure 1.

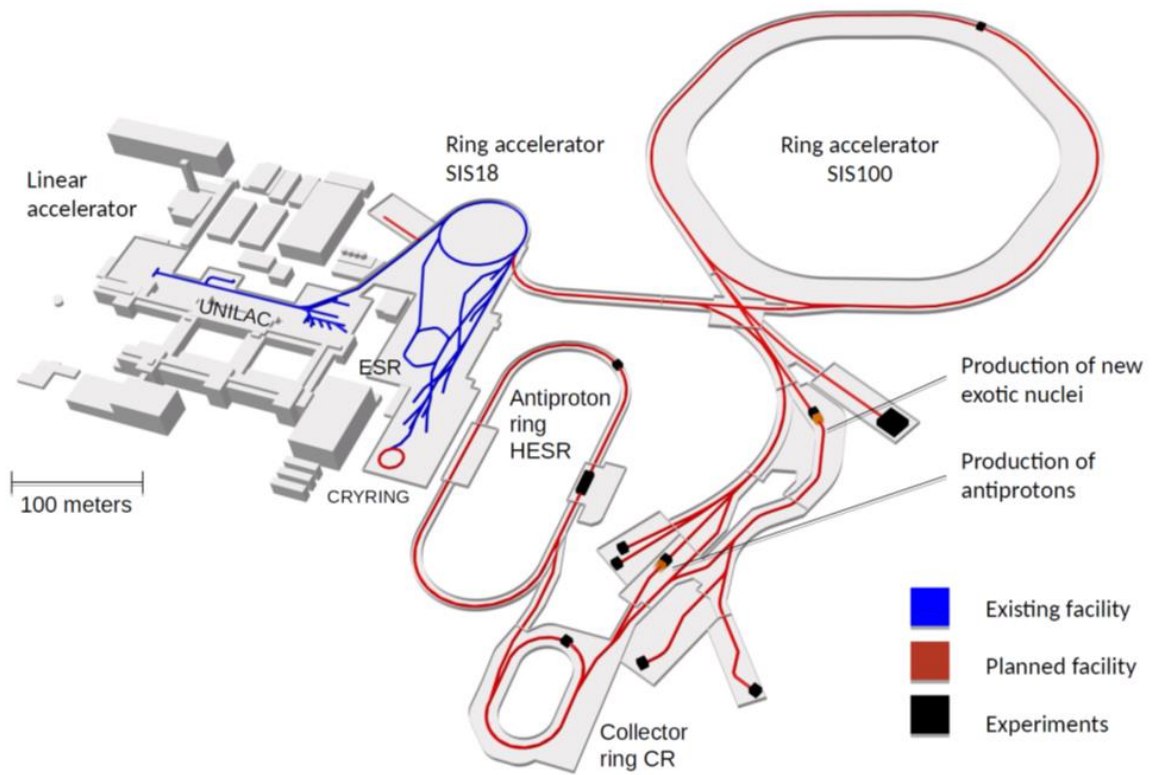

Fig. 1: Layout of the future Facility for Antiproton and Ion Research [1].

## 2. The research program of the CBM experiment

The scientific goal of the CBM experiment is to find answers to the following fundamental questions:
- What is the high-density equation-of-state of nuclear matter, which is relevant for our understanding of supernova, the structure of neutron stars, and the dynamics of neutron star mergers?
- What are the relevant degrees of freedom at high densities? Is there a phase transition from hadronic to quark-gluon matter, a region of phase coexistence, and a critical point? Do exotic QCD phases like quarkyonic matter exist?
- Can we find experimental evidence for the restoration of chiral symmetry, in order to shed light on the generation of hadron masses?
- How far can we extend the chart of nuclei towards the third (strange) dimension by producing single and double hypernuclei? Which role do hyperons play in the core of neutron stars?

These questions will be addressed by measuring the following observables:
- The equation-of-state can be studied by measuring (i) the collective flow of identified particles, which is generated by the density gradient of the early fireball, and (ii) by multi-strange hyperons, which are preferentially produced in the dense phase of the fireball via sequential collisions.
- The existence of a phase transition from hadronic to partonic matter is expected to be reflected in the following observables: (i) the excitation function of multi-strange hyperons, which are driven into equilibrium at the phase boundary; (ii) the excitation function of the invariant mass spectra of lepton pairs which reflect the fireball temperature, and, hence, may reveal a caloric curve and a first-order phase transition; (iii) the excitation function of higher-order event-by-event fluctuations of conserved quantities such as strangeness, charge, and baryon number may be are expected to occur in the vicinity of the critical point ("critical opalescence").

- Modifications of hadron properties in dense baryonic matter and the onset of chiral symmetry restoration will affect the invariant-mass spectra of di-leptons, which will be measured both in the electron and the muon channel with unprecedented precision.
- The discovery of new (double-Λ) hypernuclei, and the measurement of their life time will provide information on the hyperon-nucleon and hyperon-hyperon interaction, which will shed light on the hyperon puzzle in neutron stars.

An exhaustive review of the physics of hot and dense QCD matter is presented in the CBM Physics Book [2]. A more recent overview on the CBM research program can be found in [3].

## 3. The CBM detector setup

The research program on dense QCD matter at FAIR will be performed by the experiments CBM and HADES which are shown in figure 2. The two setups will be operated alternatively. The HADES detector, with its large polar angle acceptance ranging from 18 to 85 degrees, is well suited for reference measurements with proton beams and heavy-ion collision systems with moderate particle multiplicities, such as Ni+Ni or Ag+Ag at low SIS100 energies. With the HADES detector, electron-positron pairs and hadrons including multi-strange hyperons can be reconstructed.

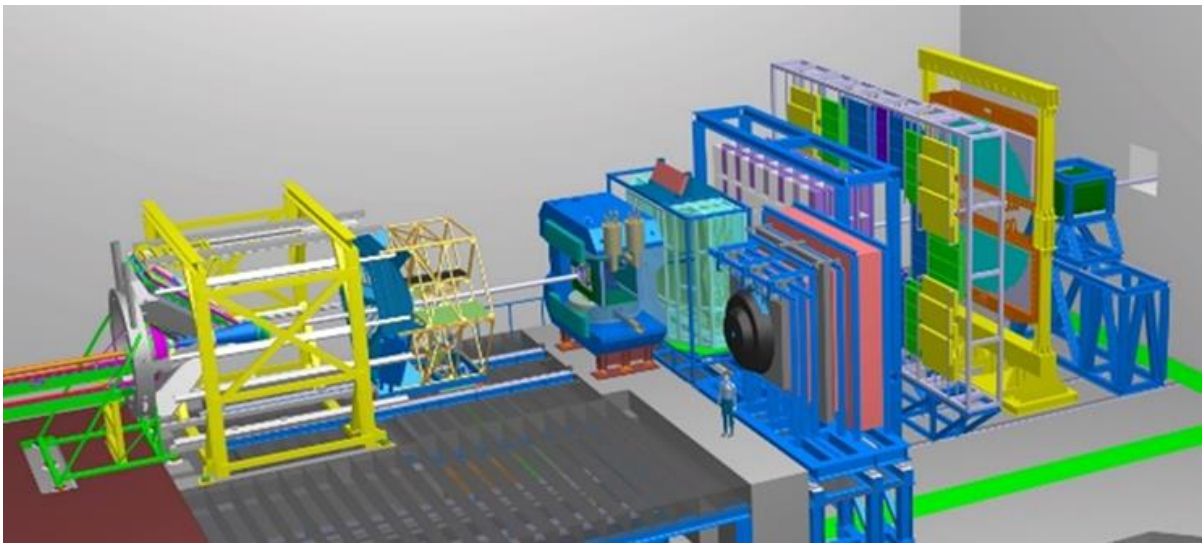

Fig. 2: The HADES detector (left) and the CBM experimental setup (right) (see text).

The CBM setup is a fixed target experiment which will be capable to measure hadrons, electrons and muons in heavy-ion collisions over the full FAIR beam energy range. In a central collision of two gold nuclei at FAIR energies, about 700 charged particles are emitted. The tracks of these particles will be measured by a Silicon Tracking System consisting of 8 layers of double-sided micro-strip sensors located in a magnetic field of a superconducting dipole magnet. The identification of hadrons requires an additional time-of-flight measurements, which is performed by a wall of Multi-Gap Resistive Plate Chambers (MRPC) with an active area of about 120 m$^2$ located about 7 m downstream the target. A Micro-Vertex-Detector (MVD), consisting of four layers of silicon monolithic active pixel sensors located in front of the STS, provides high-precision information of secondary decay vertices of short-lived particles like of open charm mesons.

The identification of electrons and positrons is performed by a Ring Imaging Cherenkov (RICH) detector and a Transition Radiation Detector (TRD). Muons will measured by a Muon Chamber (MuCh) system which consists of Gas Electron Multiplier (GEM) detector triplets sandwiched between hadron absorber plates made of graphite and iron. For muon measurements, the RICH detector will be replaced by the MuCh, and the TRD serves as tracking detector after the last hadron absorber. The reaction plane angle of the collision will be measured with the Project Spectator Detector, which is a segmented hadronic calorimeter located about 10 m downstream the target.

The data readout chain of the CBM experiment is based on free streaming frontend electronics that deliver time-stamped signals from each detector channel without trigger, i.e. without event correlation. Based on the time and position information of the detector signals, first the tracks, and then the particles, and finally the events are reconstructed and selected online by high-speed algorithms tuned to run on modern multi-core CPU architectures, which are located in the high-performance computing farm of GSI, the GreenIT cube.

Detailed information on the various hardware and software developments towards the realization of the CBM experiment is presented in the CBM Progress Report 2018 [4].

### 3.1 The superconducting magnet

The large-aperture superconducting dipole magnet hosts the target and the Silicon Tracking System, and provides the magnetic field needed for the momentum determination of the charged particles produced in the collisions. The magnet has been jointly designed by experts from GSI Darmstadt, the Institute for Nuclear Research in Dubna, Russia, and from the Budker Institute for Nuclear Physics (BINP) in Novosibirsk, Russia, Presently, the magnet is under construction at BINP. The superconducting cable has been produced, the yoke is being machined, the coils and the cryogenic system has been designed. The cable has a Cu/NbTi ratio of 7. The coils are surrounded by a copper case, and are indirectly cooled by liquid Helium flowing in a tube embedded in the copper case. The cooling circuit is of thermo-syphon type. The yoke of the magnet has a weight of 120 tons, the aperture is 144 cm vertically and 300 cm horizontally. The maximum field is 1 T, the field integral is 1 Tm over a distance of ± 0.5 m around the center. The magnet is depicted in figure 3.

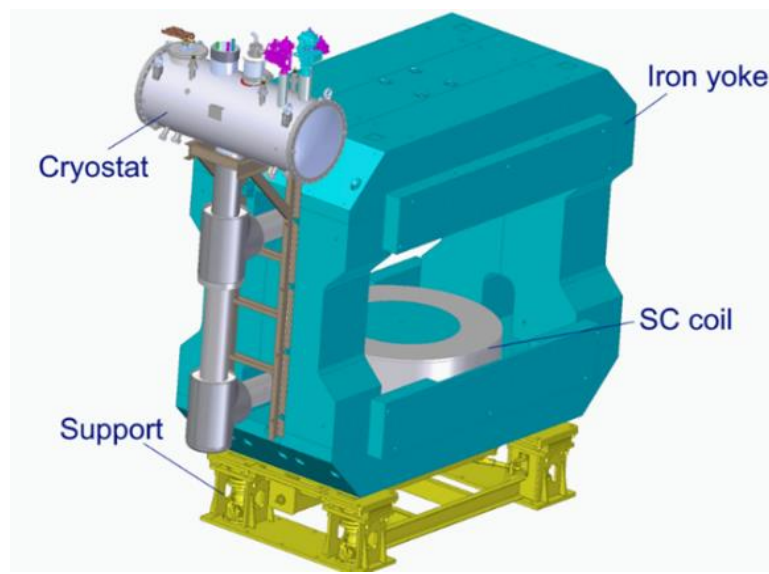

Fig. 3: The superconducting CBM magnet. The weight of the yoke is 120 tons, the gap has a height of 144 cm, and a width of 300 cm.

## 3.2 The Micro-Vertex Detector

The first detector system after the target is the Micro-Vertex Detector (MVD) [6]. The MVD has several tasks: (i) to measure the decay vertices of short-lived particles like mesons containing charm quarks, (ii) to reconstruct tracks of instable particles which one neutral daughter such as Σ hyperons which then can be identified via the missing mass method, and (iii) to measure charged particles with very small relative emission angles such a as electron-positron pairs from gamma conversion in the target. These "close pairs" have to be rejected, in order to reduce the background for lepton measurements. Moreover, the MVD supplements the STS detector with respect to primary vertex reconstruction, and, hence, suppression of fake tracks. The MVD comprises 4 stations, located at distances of 5, 10, 15, and 20 cm downstream the target, and is composed of Monolithic Active Pixel Sensors (MAPS). The engineering design of the MVD including the cooling system is illustrated in figure 4.

The future CBM pixel sensor type MIMOSIS represents a derivative of the ALICE pixel sensor ALPIDE. Due to the CBM fixed target geometry and the extremely high interaction rates, the CBM sensor has to be further optimized. The current MIMOSIS-0 sensor version has been successfully tested after application of an ionizing dose of 20 Mrad, and a neutron equivalent fluence of up to $10^{14}$ $n_{eq}$ cm$^{-2}$, which is expected to be the life dose of the CBM experiment. The first full-size CBM-pixel sensor MIMOSIS-1 is currently being produced. One MIMOSIS chip will comprise 1024 columns of 504 pixels each, with a pixel size of 26.88 x 30.24 μm$^2$. More detailed information on the MVD is given in [4,5]. The MVD is being jointly developed by the University of Frankfurt and IPHC Strasbourg.

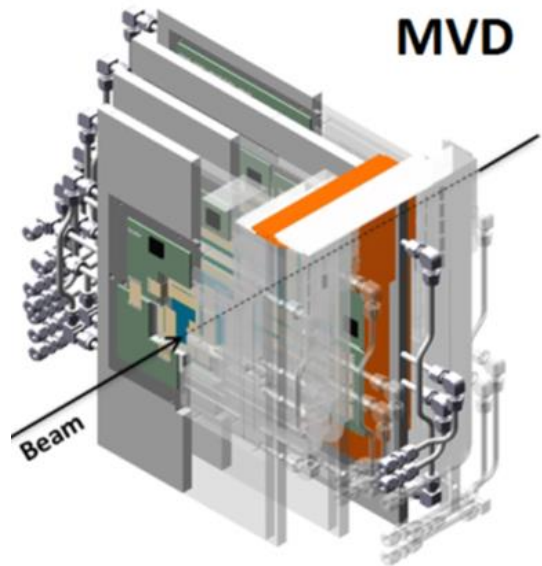

Fig. 4: Engineering design of the Micro-Vertex Detector

## 3.3 The Silicon tracking system

The task of the Silicon Tracking System (STS) is to provide track reconstruction and momentum determination of charged particles. The multiplicity of charged particles is up to 700 per event within the detector acceptance. The STS consists of 8 tracking stations of silicon double-sided micro-strip sensors covering polar emission angles between 2.5° and 25°. The stations are located downstream of the target at distances between 30 cm and 100 cm inside the magnetic dipole field. The required momentum resolution is of the order of Δp/p = 1-2%. This performance can only be achieved with an ultra-low material budget of the stations, imposing particular restrictions on the location of power-dissipating front-end electronics in the fiducial volume. The concept of the STS tracking is based on silicon micro-strip sensors mounted onto lightweight mechanical support ladders. The sensors will be read out through multi-line micro-cables with fast electronics at the periphery of the stations where cooling lines and other infrastructure can be placed. The micro-strip sensors will be double-sided with a stereo angle of 7.5°, a strip pitch of 58 μm, strip lengths between 20 and 120 mm, and a thickness of 300 μm of silicon. The micro-cables will be built from sandwiched polyimide-Aluminum layers of several 10 μm thickness.

The STS consists of 890 modules, each comprising a double-sided silicon sensor with 2048 strips, 32 micro-cables each with 64 lines, 2 Front-End-Board each carrying 8 ASICs with 128 channels each. A photo of a module is shown in figure 5. The readout ASIC for the STS and MUCH detectors is a 128 channel integrated circuit for time (14-bit timestamp and 3.125 ns resolution) and energy (5-bit ADC) measurements. This ASIC has been designed at AGH University of Science and Technology in Krakow, Poland [6].

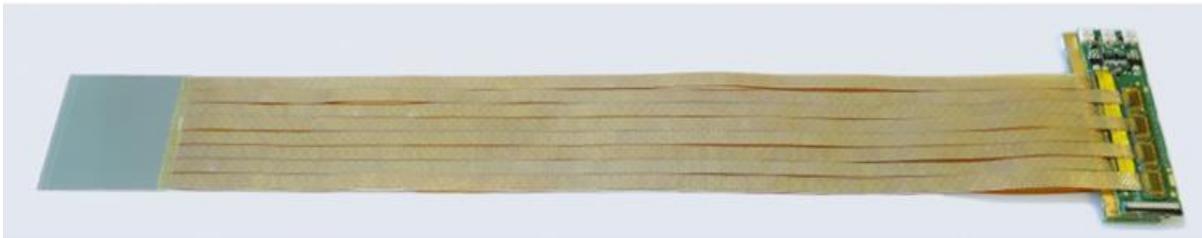

Fig. 5: A STS module comprising a sensor, micro-cables, Front-End-Boards, and ASICs (see text).

The STS is operated in a thermal enclosure that keeps the sensors at a temperature of about -10°C, in order to limit radiation induced leakage currents. The technical challenge is to remove the heat dissipated in the close-by read-out electronics, which amounts to about 40 kW. For this purpose, a $CO_2$ bi-phase cooling system, and, alternatively, a mono-phase NOVEC cooling are under investigation. Figure 6 depicts a sketch of the STS including its mainframe, which will position the detector in the magnet gap. The STS is being realized in a collaboration of GSI and FAIR Darmstadt, Joint Institute of Nuclear Research (JINR) in Dubna, University of Tübingen, Karlsruhe Institute of Technology, Jagiellonian University and AGH UST Krakow, Kiev Institute for Nuclear Research, and Warsaw University of Technology. Additional information on the STS is presented in [4,7].

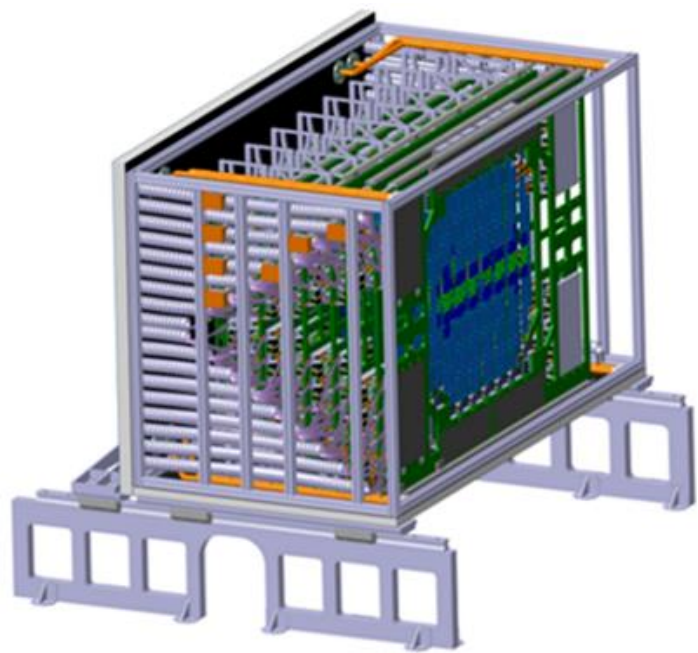

Fig. 6: The CBM Silicon Tracking System

## 3.4 The Time-of-Flight wall

The identification of the produced particles requires not only the determination of their momentum with the STS, but also the measurement of their time-of-flight (TOF). This is performed with an array of Multi-gap Resistive Plate Chambers (MRPCs). This detector covers an active area of about 120 m$^2$ and is located about 7 m downstream of the target. The required time resolution for the whole detector system is of the order of 80 ps. The detector arrangement as sketched in figure 7 comprises five different MPRC module types, which differ in size, granularity, and rate capability, according to the expected particle flux rate for central Au+Au collisions at high FAIR energies. All modules have a width of 32 cm, and are divided in 32 vertical strips with a length according to the module height. The modules in the outer region of the wall, where the hit rate is expected to be below 1 kHz/cm$^2$, consist of standard float glass, and have a height of 50 cm (turquoise color) and 27 cm (dark green color). These modules are built by the University of Science and Technology of China (USTC) in Hefei. The modules in the inner sector, where the hit rates are expected to be much higher, are made of low-resistivity glass. The modules colored in light green have a height of 27 cm and will be operated at rates up to 10 kHz/cm$^2$. These modules are built by the Tsinghua University (THU) in Beijing, China. The modules colored yellow and red consist also of low resistivity glass, and vary in length between 12, 6, and 4 cm, in order to accommodate rates up to 20 kHz/cm$^2$. These modules are built and have been tested by the Horia Hulubei National Institute of Physics and Nuclear Engineering (HH NIPNE) in Bucharest, Romania. An update of the development activities can be found in [4,8].

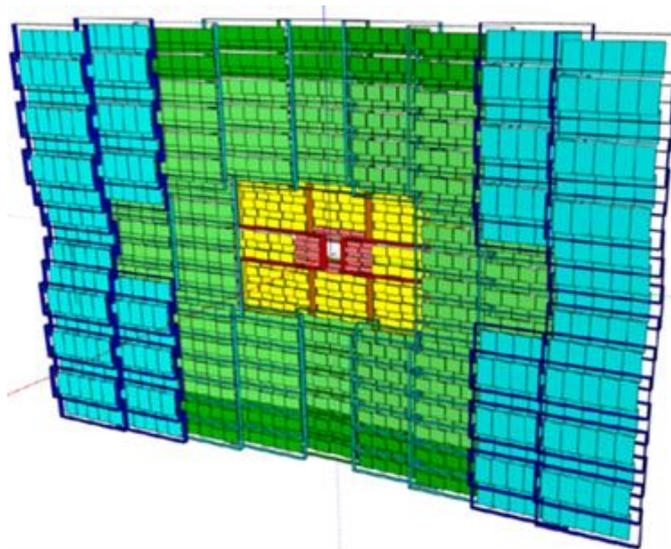

Fig. 7: MRPC module arrangement for the CBM TOF wall. The colors refer to counter types which differ in size, granularity, and resistivity of the glass electrodes, in order to accommodate the different requirements for rate capability (see text).

Meanwhile, the pre mass-production of the 3rd generation "MRPC3" counters for the TOF FAIR phase 0 program is finished. More than 70 high-rate MRPC3a with low-resistivity glass electrodes have been built. Out of this batch, 48 counters together with 60 MRPC3b counters made of float glass have been installed in a wheel-like arrangement as end-cap TOF detectors in the STAR experiment at RHIC/BNL, contributing to the STAR upgrade program. A photo of this detector system is shown in the left panel of figure 8. The counters were read-out by the prototype version of the CBM free-streaming data acquisition. The performance of the end-cap TOF concerning particle identification is illustrated in the right panel of figure 8. It turns out, that the system time resolution is about 85 ps, which allows to separate kaons from pions up to momenta of 1.8 GeV/c. The CBM TOF project is realized by a collaboration between THU Beijing, NIPNE Bucharest, GSI Darmstadt, TU Darmstadt, University Frankfurt, USTC Hefei, University Heidelberg, ITEP Moscow, HZDR Rossendorf, and CCNU Wuhan.

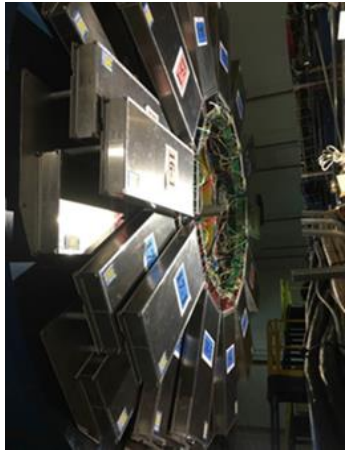 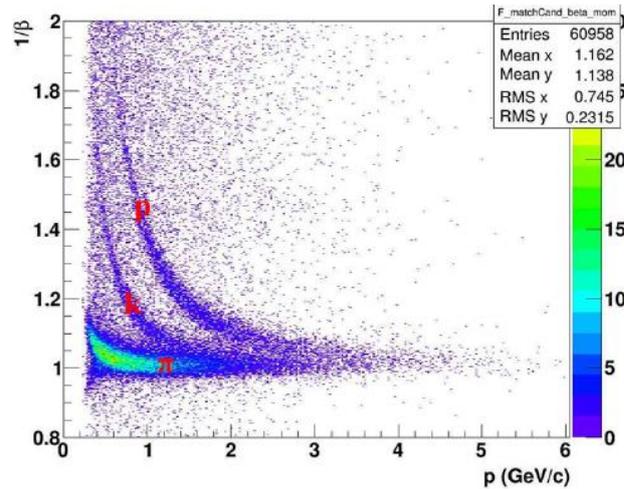

Fig. 8: Left panel: Photo of the STAR end-cap TOF system comprising 108 CBM MRPC modules. Right panel: Particle separation via end-cap TOF measurements in the STAR experiment in the plane inverse velocity versus momentum of the particles [4].

**3.5 The Ring Imaging Cherenkov detector**

The task of the Ring Imaging CHerenkov (RICH) detector is to provide the identification of electrons and suppression of pions. The RICH detector comprises a $CO_2$ gas radiator, two large focusing mirrors, and two photon cameras consisting of 1100 Multi-Anode Photo Multiplier Tubes (MAPMTs) with 64 channels each. A sketch of the RICH detector is shown in figure 9. Each mirror comprises 8 x 10 tiles, each of size 40 x 40 cm$^2$, resulting in a total mirror size of about two times 13 m$^2$. The tiles are made of glass, and have to be adjusted with high precision.

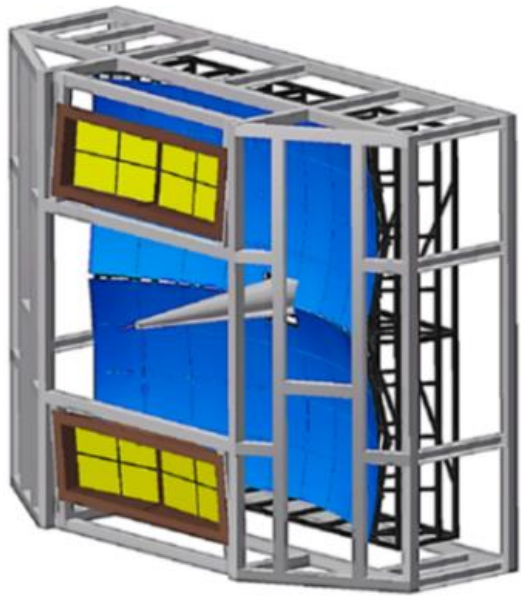

Fig. 9. Sketch of the CBM RICH detector. Each of the two mirrors (shown in blue) consists of 80 glass tiles with a total size of about 13 m$^2$. The two photon detectors (shown in yellow) comprise 1100 multi-anode photo multiplier tubes with 64 channels each. The radiator gas is $CO_2$.

The MAPMTs of type H12700 from Hamamatsu have been delivered and tested. Out of 1100 MAPMTs, 428 have been installed in the HADES experiment at SIS18 of GSI, to replace the old RICH photon detector. A photo of the new HADES RICH is shown in the left panel of figure 10. The right panel of figure 10 depicts an online event display measured by HADES RICH in an Ag+Ag collision at 1.65A GeV. The hit pattern contains two rings produced by an electron-positron pair, each ring comprising about 16 pixels depending on the cuts. The CBM and HADES RICH detector readout electronics is based on the TDC Readout Board (TRB) with FPGA based time-to-digital converters (TDCs). The CBM RICH detector is realized by a collaboration of the University of Giessen, University of Wuppertal, the Petersburg Nuclear Physics Institute Gatchina, and GSI Darmstadt. More information on the actual status of the RICH developments is presented in [4].

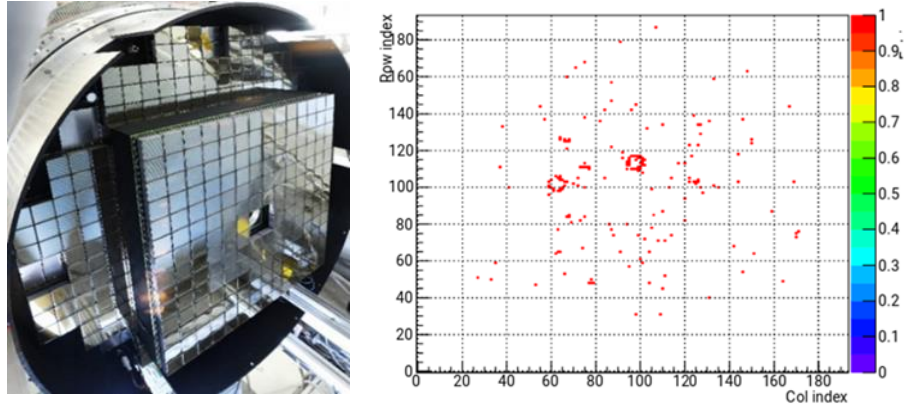

Fig. 10: Left: The new HADES photon detector build of 428 H12700 MAPMTs as developed for the CBM RICH. Right: Online event display from the HADES RICH with no timing cuts applied yet, measured in Ag+Ag collision at an energy of 1.65A GeV (see text).

### 3.6 The Transition Radiation Detector

The main task of the Transition Radiation detector (TRD) is to identify electrons above momenta of 1 GeV/c and thus to extend the electron identification capabilities of the Ring Image CHerenkov (RICH) detector above momenta of $p \approx 5$ GeV/c. This identification has to be achieved with a pion suppression factor of about 20 at 90% electron efficiency, in order to allow for a measurement of dielectrons in the mass range from below the ρ and ω masses to beyond the J/ψ mass with a good signal-to-background ratio. Due to its capability of identifying charged particles via their specific energy loss, the TRD in addition will provide valuable information for the measurement of nuclear fragments. This is in particular important for the separation of, e.g, deuterons and $^4$He, which cannot be achieved using a time-of-flight measurement alone.

The baseline design of the TRD foresees one station composed of four detector layers as illustrated in figure 11. It will be positioned between the RICH and the Time-Of-Flight (TOF) detector, and thus allows to reduce the background in the TOF resulting from track mismatches by providing additional position information for high precision tracking between Silicon Tracking System (STS) and TOF. In the muon configuration of CBM the TRD will also be used as tracking station behind the last absorber of the MUCH detector.

The TRD is designed such that the entire system can be built with modules of only two different sizes, i.e. with outer dimensions of $57 \times 57$ cm$^2$ and $99 \times 99$ cm$^2$, which have four different granularities. Each detector layer comprises 10 small modules with a pad size of 1.2 cm$^2$ in the innermost region, then 24 small modules with 4.6 cm$^2$ pads, 8 large modules with 2.7 cm$^2$ pads, and in the outer region 12 large modules with 8 cm$^2$ pads. The total active detector area amounts to about 114m$^2$. The TRD comprises 216 modules, which are Xe/CO$_2$ based Multi-Wire Proportional Counter (MWPC) detectors in combination with PolyEthylene (PE) foam foils as radiator. The default MWPC design is composed of a symmetric amplification area of 3.5 + 3.5 mm thickness, complemented by a 5 mm drift region to enhance the TR-photon absorption probability in the active gas volume. This geometry provides also efficient and fast signal creation, as well as readout, with timescales below 250 ns per charged particle track. The performance of the detector is optimized by reducing the material budget between radiator and gas volume to a minimum.

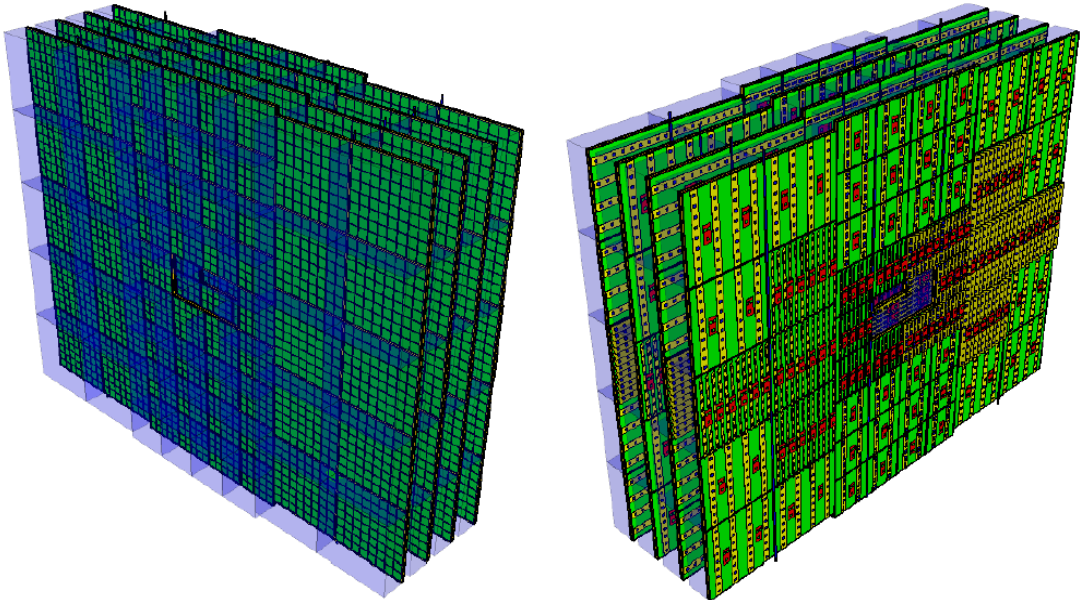

Fig. 11: Geometry of the TRD station with 4 detector layers as implemented in the CBM simulation framework. Visible are the read-out chambers (ROCs) with the radiator boxes in the front view (left), while the rear view (right) shows the back panels of the ROCs together with the front-end electronics

The TRD readout architecture is based on a mixed-signal readout ASIC with 32 input channels developed explicitly for the readout of the TRD of CBM. This ASIC will be operated on the Front-End-Boards (FEBs), directly mounted on the back of MWPCs. The subsequent read-out chain is similar for most of the sub-detector systems, and is described in chapter 3.9. More information on the TRD can be found in [4] and in the recently finalized Technical Design Report [9]. The TRD is realized by NIPNE Bucharest, University Frankfurt, University Heidelberg, and the University Münster.

**3.7 The Muon Detection System**

The measurement of muons in heavy-ion collisions at energies of 2 -10A GeV is a particular challenge because of the soft muon momenta and the high particle multiplicities. Muon pairs have been measured in Indium-Indium collisions at 158A GeV by the NA60 experiment at the CERN-SPS, which was the lowest beam energy so far [10]. In order to identify soft muon pairs in a large combinatorial background, the CBM we will use a set of instrumented hadron absorbers with tracking detector triplets in-between. The numbers of installed absorbers and detector stations will depend on the beam energy. For beam energies above 4A GeV, the Muon Chamber (MuCh) system comprises a 60 cm thick carbon/concrete absorber placed right after the magnet, followed by iron absorbers with thicknesses on 20 cm, 20 cm, 30 cm, and 100 cm. This configuration is depicted in figure 12.

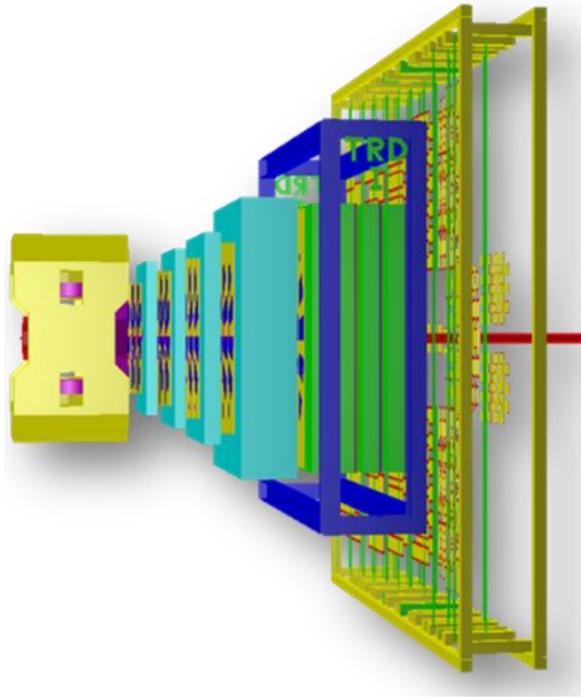

Fig. 12: The CBM detector setup used for muon experiments. From left to right: Magnet (yellow) hosting the STS, the MuCh with 5 absorbers (one pink and four turquoise) sandwiching 2 triplets of GEM detectors and 2 triplets of RPC detectors, the TRD (green radiators and multi-wire chambers with blue support structure) after MuCh, and finally the TOF wall (yellow).

Charged particles passing the absorbers are tracked by detector triplets after each absorber. The technology of the gaseous muon tracking detectors is matched to the hit density and rate: behind the first and second hadron absorber (particle density up to 500 kHz/cm$^2$ we will install Gas Electron Multiplier (GEM) Detectors. Prototype GEM detectors with single-mask foils have been successfully tested with particle beams at CERN. The GEM stations are of circular shape with the beam pipe in the centre. The first GEM station consists of 16 trapezoidal detector modules with a length 80 cm each, whereas the second, larger station comprises 20 modules with a length of 97 cm. The readout consists of pads with progressively increasing pad sizes from $\approx$ 3mm to $\approx$ 17mm. The production of the modules has already started. Further downstream at stations 3 and 4, where the hit density is reduced to about 15 kHz/cm$^2$ and 4 kHz/cm$^2$, respectively, the use of single-gap RPC detectors with electrodes from low-resistivity Bakelite is under investigation. Additional information on the MuCh hardware developments are presented in [4].

The data from the GEM detectors are readout via the same ASIC as for the Silicon detector (see chapter 3.3), and are further processed by the CBM data acquisition system (see chapter 3.9). The MuCh project is realized by 13 Indian institutions under the leadership of VECC and Bose Institute in Kolkata, and by PNPI Gatchina.

### 3.8 The Projectile Spectator Detector

The Projectile Spectator Detector (PSD) of the CBM experiment is a compensating lead-scintillator calorimeter designed to measure the energy distribution of the forward going projectile nucleons and nuclei fragments (spectators) produced close to the beam rapidity in high intensity beams at FAIR SIS100 up to $10^6$ interactions/s. This information will be used to determine the orientation of the reaction plane, and the centrality of the collision. The PSD has 44 modules with the beam hole in the center. Each module consists of 60 lead/scintillator sandwiches with 4 mm thick scintillator tiles and 16 mm lead plates. All sandwiches are wrapped in 0.5 mm stainless steel sheet, and tied together in one block with a length of about 120 cm (5.6 nuclear interaction lengths). The transverse size of the modules is 20x20 cm$^2$, and the weight of each module is about 500 kg. A very similar hadron calorimeter is presently used at the NA61 experiment at the CERN-SPS, where its response has been studied at proton energies in the range of 20 - 150 GeV. Moreover, such a calorimeter will be installed at the BM@N experiment at the Nuclotron at JINR, and also is being developed for the MPD experiment at NICA. A drawing of the future CBM-PSD including its support structure is presented in figure 13.

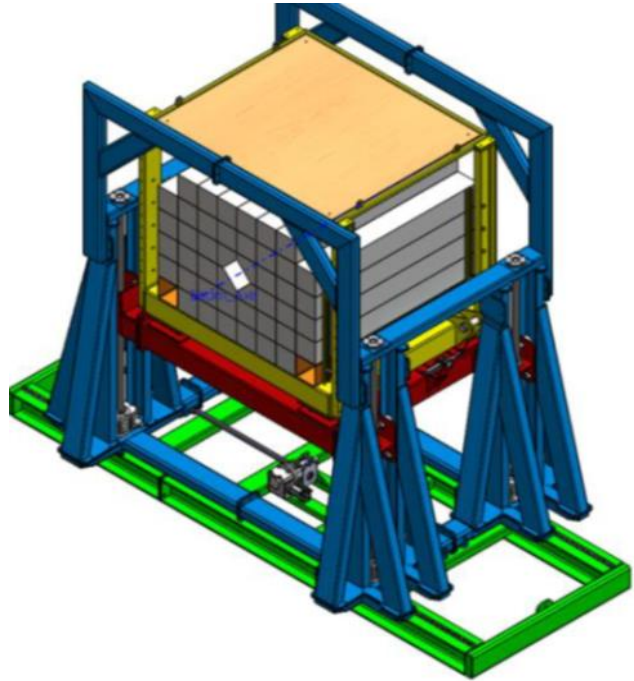

Fig. 13: The Projectile-Spectator Detector (PSD) for the CBM experiment.

The CBM-PSD consists of 44 modules with 440 readout channels. Each channel will be connected with one Silicon Photomultiplier (Multi-Pixel Photon Counter MPPC Hamamatsu S12572-010P) which produces a signal with a width of $\approx$ 50 ns. The readout electronics is based on the developments for the PANDA and HADES experiments at FAIR/GSI. The PSD detector will be built by INR Moscow, NPI CAS Rez, Czech Tech. University Prague, and Tech. University Darmstadt. Further information on the CBM-PSD is available at [4].

**3.9 The CBM Data readout and acquisition system**

The CBM experiment at FAIR will measure relativistic nucleus-nucleus collisions with collision rates up to 10 MHz leading to data rates up to 1TB per second. To achieve the required performance a free-streaming, triggerless data acquisition system is being developed. All detector signals are readout by ultra-fast and radiation-tolerant ASICs as front-end chips, which mark each hit with a time stamp, and send it to CERN GBTx-based radiation-tolerant data aggregation units. Further down-stream, the data streams are handled by Common Readout Interface (CRI) boards containing powerful FPGAs, and are forwarded via a PCIe based FPGA board to the entry node of a large-scale computer farm, the First-Level Event Selector (FLES). Finally, the data are transmitted via a high-speed optical connection (InfiniBand) to the FLES processing nodes, which are located in the high-performance computing center of GSI, the GreenIt cube. Upon reception in a FLES processing node, the micro-slices originating from all active subsystems, i.e the detector hits with time stamps collected over about 100 μs, are grouped into larger time-slices of about 100 ms, containing information from all detector systems. These time-slices are then used for the online data reconstruction and selection. A sketch of the CBM data readout and processing chain is depicted in figure 14. The high-performance readout hardware requires high-performance firmware for the FPGA layers, and a fast and highly parallelized software for online track and event reconstruction and analysis. The CBM data readout and reconstruction chain is developed by groups from GSI Darmstadt, Frankfurt Institute for Advanced Studies (FIAS), University Heidelberg, AGH University of Science and Technology in Krakow, University Frankfurt, Karlsruhe Institute of Technology, Indian Institute of Technology Kharagpur, Warsaw University of Technology.

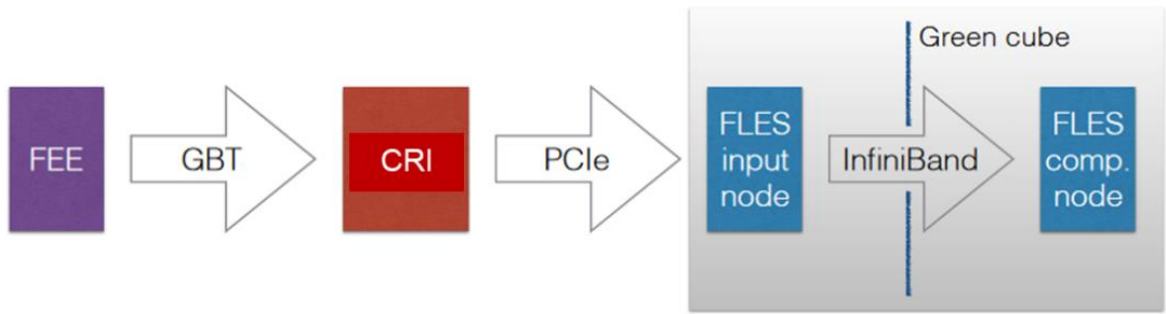

Fig. 14: The CBM data readout and acquisition chain (see text)

### 3.10 The miniCBM Data experiment at SIS18/GSI

The primary purpose of the miniCBM (mCBM) experiment at SIS18 is to develop, commission and optimize (i) the free-streaming data acquisition system including the data transport to a high performance computer farm inside the GreenIT Cube, (ii) the online track and event reconstruction and event selection algorithms and (iii) the offline data analysis as well as the controls software package. The mCBM setup will comprise final prototypes and pre-series components of all CBM detector subsystems and their read-out systems. Hence, the setup offers additional high-rate detector tests in nucleus-nucleus collisions under realistic experiment conditions up to interaction rates of 10MHz.

As illustrated in figure 15, the mCBM test-setup is positioned downstream a solid target under a polar angle of about 25° with respect to the primary beam. The setup does not feature a magnetic field, and, therefore, will measure charged particles produced in nucleus-nucleus collisions traversing the detector stations following straight trajectories. In the final configuration, the mCBM tracking system comprises nine full-size prototype detectors: two STS stations, three MuCh-GEM detectors, and four TRD (mTRD) stations, which provide redundant position information and allow to perform tracklet searches. Moreover, the mCBM setup includes a high-resolution time-of-flight system consisting of a fast and 8-fold segmented diamond counter for time-zero determination in front of the target as well as a TOF stop wall. An aerogel type RICH detector will be placed behind the mTOF detector, and will deliver a second measurement of the particle velocity in a selected acceptance window. A small electromagnetic calorimeter will also be mounted behind the mTOF. In addition, a prototype PSD module will be positioned directly under the beam pipe, tilted relative to the beam axis while pointing to the target.

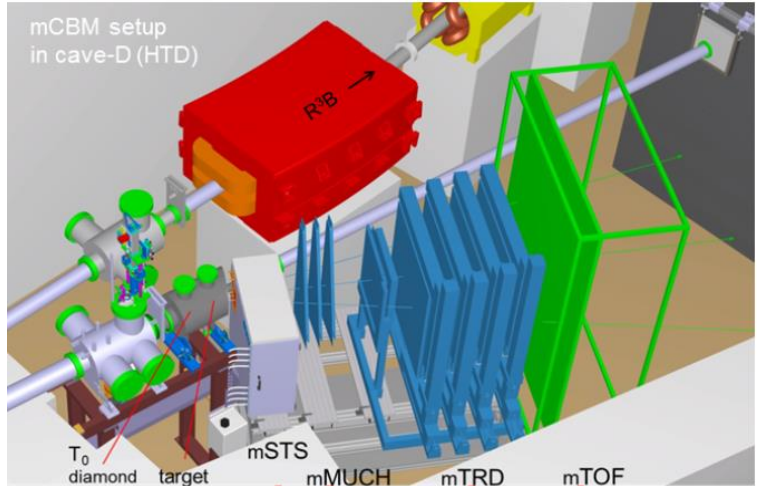

Fig. 15: Sketch of the mCBM setup installed at SIS18/GSI (see text).

The detector systems are equipped with prototype versions of the self-triggered front-end electronics, which are read out by the data acquisition chain similar to the final one as described in the previous chapter. This prototype version of the free-streaming data acquisition system is used to aggregate the data, and to transport it to an online compute farm for data reconstruction and selection in real time. The successful operation of the data read-out and acquisition hardware requires the implementation of high-performance firmware for the FPGA layers, as well as software for a fast and highly parallel on-line track and event reconstruction and selection. The first commissioning of the mCBM setup as shown in figure 16 has been performed in December 2018. Detailed information on the setup can be found in [4].

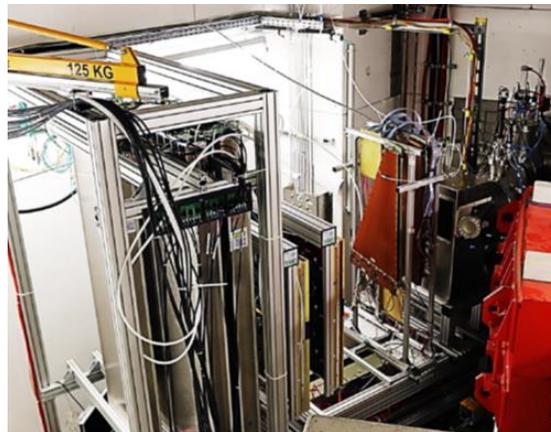

Fig. 16: photo of the mCBM test setup at GSI-SIS18 as used in the first commissioning run in December 2018 (see text).

**4. Physics performance simulations**

**4.1 Track reconstruction using STS information**

The task of the STS is to provide information which allows to reconstruct the tracks of the charged particles produced in heavy-ion collisions. The left panel of figure 17 illustrates a central collision of two gold nuclei at a beam kinetic energy of 10A GeV simulated with the UrQMD event generator [11]. The produced particles are transported through the STS using the GEANT3 code [12]. The charged particles are deflected by the magnetic dipole field. In addition to the charged hadrons produced in the collision, also delta-electrons are generated by the beam ion traversing the target. The charged particle trajectories are reconstructed using a Cellular Automaton (CA) algorithm, based on the particle hits measured by the STS [13,14]. The result is shown in the right panel of figure 17. The track reconstruction efficiency for primary tracks reaches values of more than 95% for particle momenta above 0.5 GeV/c as shown in the left panel of figure 18. For secondary tracks the reconstruction efficiency is still above 90% for momenta larger than about 1 GeV/c. The resulting momentum resolution for primary particles decreases from 2% for very low momenta to about 1.5% for momenta above 2 GeV/c, while for secondary particles it stays more or less at a level of 2% as illustrated in the right panel of figure 18.

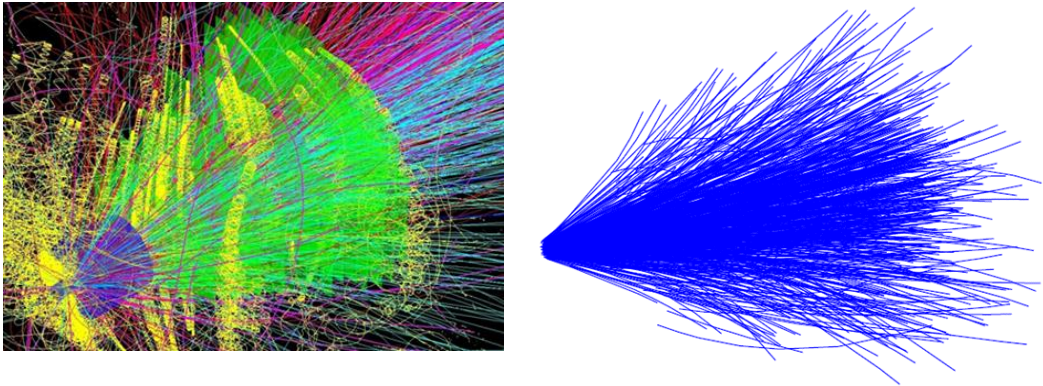

Fig. 17: Left: Simulation of trajectories of charged particles produced in central Au+Au collision at 10A GeV/c generated with the UrQMD code and transported through the STS using the GEANT3 generator. Right: reconstructed tracks using a Cellular Automaton algorithm (see text).

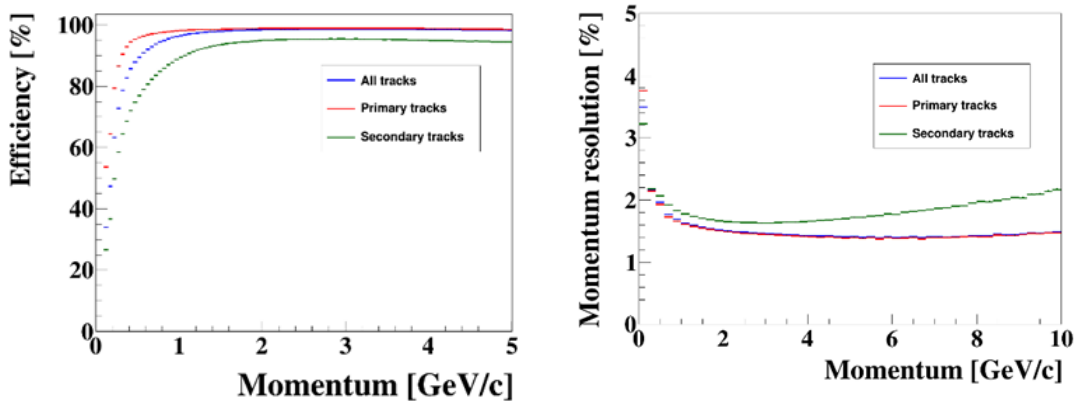

Fig.18: Track reconstruction efficiency (left panel) and momentum resolution (right panel) for primary and secondary charged particles produced in central Au+Au collisions at 10A GeV/c (see text).

### 4.2 Hadron identification using STS + TOF

The particle identification capability of the CBM TOF detector system in combination with the particle momenta measured with the STS is illustrated in figure 19, which depicts the squared mass of the charged particles as function of their momentum times charge. The particles have been simulated and reconstructed for central Au+Au collisions at 10A GeV/c. Based on the identified primary particles as shown in figure 19, also secondary particles have been reconstructed using a Kalman Filter Particle Finder [15]. The results are presented in figure 20, which depicts the invariant mass distributions of strange particles as simulated with the UrQMD event generator for central Au+Au collisions at 10A GeV/c, and reconstructed using the CBM analysis tools.

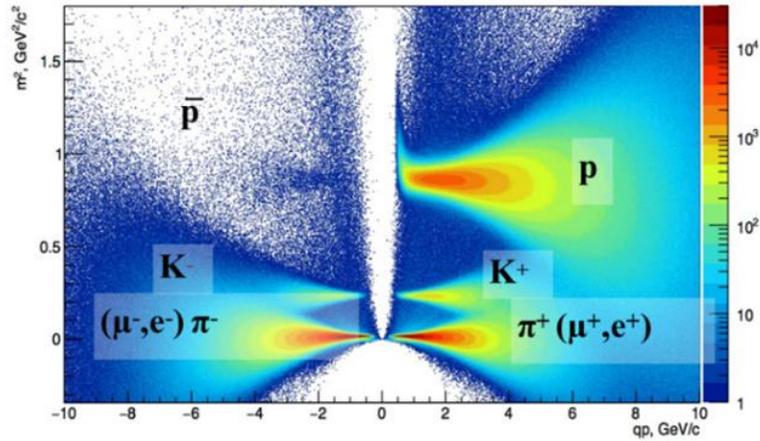

Fig. 19: Particle identification capability of the TOF detector in combination with the STS.

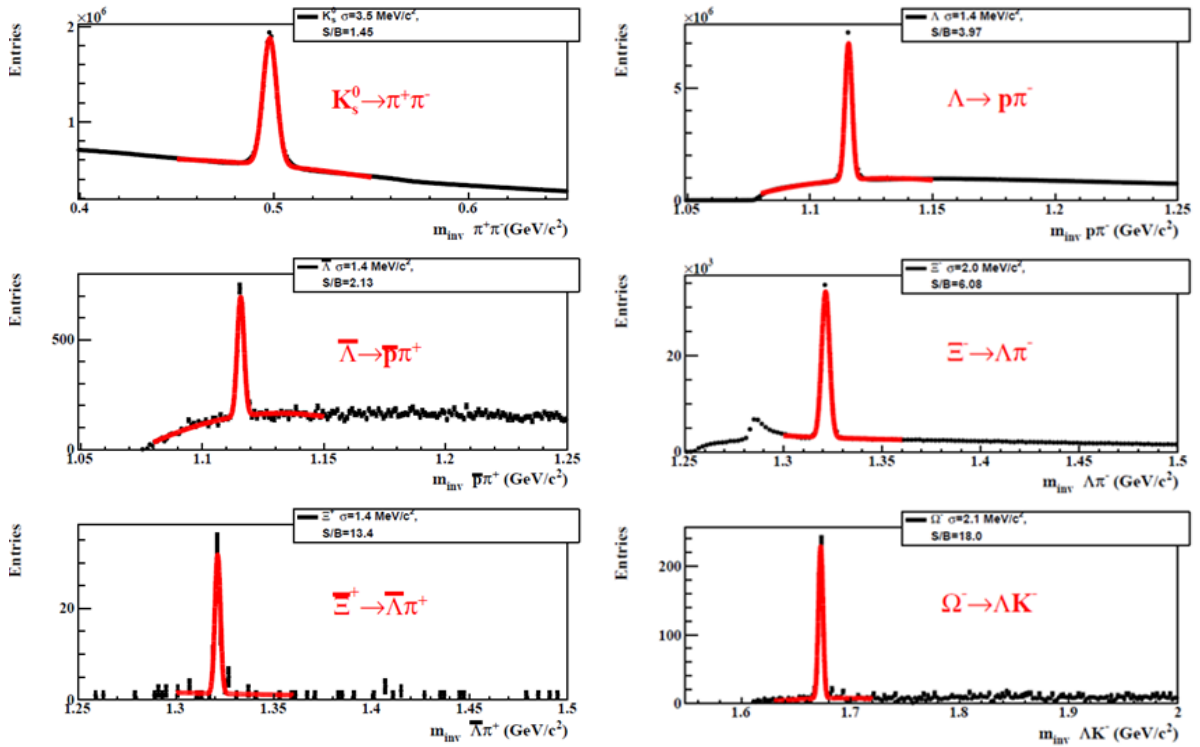

Fig. 20: Invariant mass spectra of $K^0$ mesons and (anti-) hyperons simulated and reconstructed in central Au+Au collisions at 10A GeV/c. Taken from [4].

### 4.3 Electron identification

The expected performance of the RICH has been studied for central Au+Au collisions at 10A GeV/c based on the geometry and response of the involved detector systems. The resulting Cherenkov ring-size as function of particle momentum is presented in figure 21, demonstrating the separation capability of the RICH detector for electrons and pions up to momenta 8-10 GeV/c. For pion suppression at higher particle momenta, TRD will be used. The performance of the TRD is illustrated in figure 22 by the red histogram, indicating that the pion suppression factor by the TRD reaches a value of about 50 (25) for pion momenta above about 8 GeV/c, and an electron efficiency of 80% (90%). The pion suppression factor of the RICH shown in blue lies between 700 and 1000 in the momentum range between 2 and 6 GeV/c, and then decreases rapidly to a value of about 2 at 8 GeV/c. The overall suppression factor is illustrated by the green histogram in figure 22.

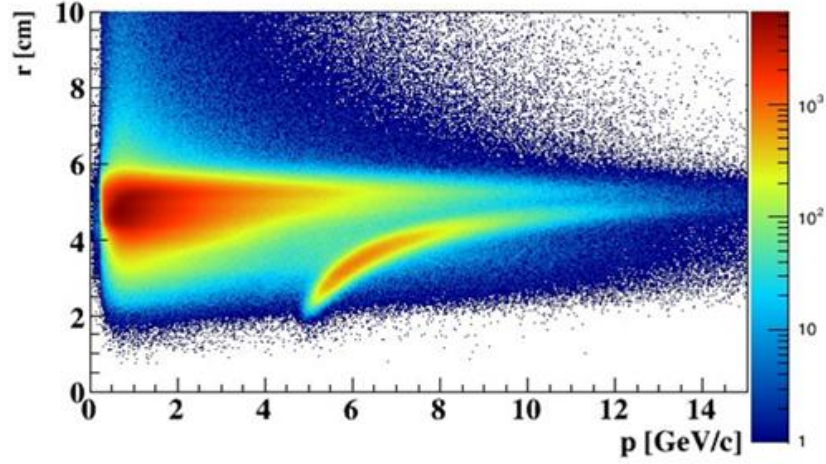

Fig. 21: Cherenkov Ring radius simulated for electrons (r = 4–6 cm) and pions as function of particle momentum for the CBM RICH detector.

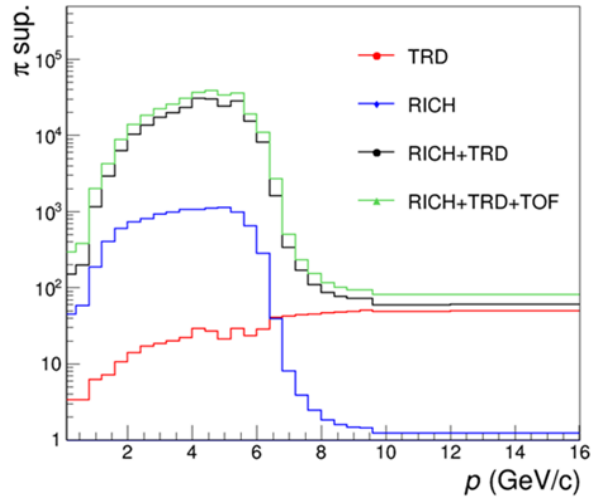

Fig.22: Simulated pion suppression factors for the RICH (blue), for the TRD (red), and for the combination of RICH+TRD (black) and RICH+TRD+TOF (green).

In addition to the detectors for pion suppression (RICH and TRD) the Silicon Tracking System is required for momentum determination of the particles. Moreover, the Micro-Vertex Detector is used to reject close electron-positron pairs, which predominantly are created by pair conversion of gamma-quanta in the target, and contribute to the background in the di-electron spectra. In order to reduce this background further, thin targets are used, for example 25 µm thick Au foils. Figure 23 depicts a di-electron invariant mass spectrum simulated for central Au+Au collisions at 8A GeV/c. The simulation has been performed in the CBM software framework with the di-electron analysis package, and using UrQMD as event generator. In each event a di-electron pair according to a thermal energy distribution was embedded. As di-electron signals either ω, φ and in-medium ρ decays into $e^+e^-$, ω-Dalitz decays, or QGP radiation was implemented [16]. Signals were scaled later according to their estimated multiplicity and branching ratio. According to the simulation, an excellent signal-to-background ratio of about 10 can be achieved up to invariant masses of 1 GeV/c$^2$. It is worthwhile to note, that because of the low material budget of the RICH and TRD the electron measurements can be performed simultaneously to the hadron measurements discussed above.

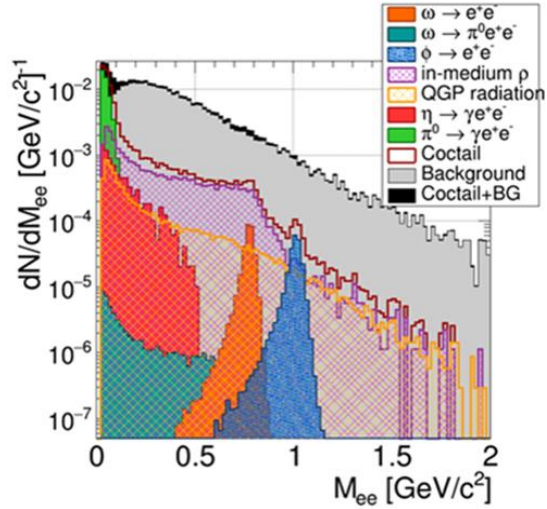

Fig. 23: Invariant electron-positron mass spectra simulated for Au+Au collisions at 8A GeV/c (see text). Taken from [4].

**4.4 Muon identification**

Figure 24 depicts the dimuon invariant mass spectrum also simulated for central Au+Au collisions at 8A GeV/c. In this case, the STS, MuCh, TRD and TOF detectors have been used. The MuCh consists of 4 tracking detector stations and 5 hadron absorber blocks, with the TRD as muon tracker after the last absorber. The magnetic field has its nominal value of 1 T, and the Au target has a thickness of 250 μm. The di-muon signals from η, ω, φ, and ρ decays, and the QGP contribution is generated like in the electron simulations. Like in the di-electron measurement, the signal-to-background ratio of the dimuon invariant mass spectrum is in the order of 10. Therefore, both the electron and muon simulations demonstrate, that the CBM detector system is very well suited to measure and identify dilepton pairs in heavy-ion collisions at SIS100 energies

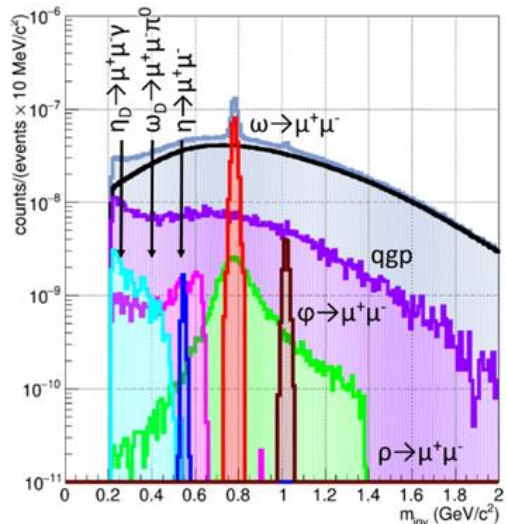

Fig. 24: Reconstructed dimuon invariant mass spectrum simulated for central Au+Au collisions with a beam momentum of 8A GeV/c (black line). The dimuon decays of low-mass vector mesons including their Dalitz-decays have been identified using Monte-Carlo information.

## 4.5 Hypernuclei reconstruction

According to model calculations, light (multi-) lambda hyper-nuclei will be abundantly produced in heavy-collisions at FSIR/SIS100 beam energies, because of the superposition of two effects: the increase of light nuclei production with decreasing beam energy, and the increase of hyperon production with increasing beam energy [17]. The performance of the CBM for the measurement of multi-strange hyperon and hyper-nuclei has been studied for central Au+Au collisions in the FAIR beam energy range using the UrQMD event generator. Hyper-nuclei will be reconstructed in the CBM experiment via their decays into charged hadrons and fragments, like $\pi$, , K, p, d, t, $^3$He, and $^4$He, which are detached from the primary vertex The primary and secondary charged particle tracks, and the secondary decay vertices have been reconstructed based on the hits delivered by the MVD and the STS. The TOF detector has been used to identify the decay products of the hyper-nuclei. The CBM performance for the identification of double-lambda hypernuclei in central Au+Au collisions at 10A GeV/c is illustrated in figure 25 for the case of $^4_{\Lambda\Lambda}$H $\rightarrow$ $^4_\Lambda$He $\pi^-$.

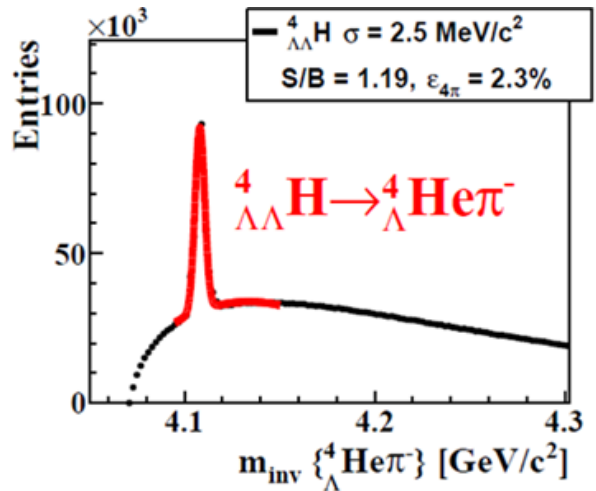

Fig. 25: The invariant-mass spectra of $^4_\Lambda$He $\pi^-$ simulated and reconstructed in central Au+Au collisions at 10A GeV/c. The red lines indicate the signal plus background fit by a polynomial plus Gaussian function.

The decay of $^4_{\Lambda\Lambda}$H $\rightarrow$ $^4_\Lambda$He $\pi^-$ is followed by a second decay $^4_\Lambda$He $\rightarrow$ $^3$He p $\pi^-$. The first step is to reconstruct the $^4_\Lambda$He via the identification of the proton, pion, and $^3$He, originating from the same decay vertex, which is detached from the primary vertex. In a second step, $^4_{\Lambda\Lambda}$H is reconstructed from its decay products. Finally, it has to be checked that $^4_{\Lambda\Lambda}$H originates from the primary vertex of the reaction. This stepwise reconstruction of a complicated topology is performed by the Kalman Filter Particle Finder package [15].

In some of the decay chains of (multi-) lambda hyper-nuclei, daughter particles like $^4$He or deuterons are formed. These particles cannot be distinguished just by measuring momentum and time-of-flight. In this case, the energy-loss information provided by the TRD helps to separate particles with the same magnetic rigidity. Figure 26 illustrates the CBM capability of light fragment identification combining information from the detector system STS, TOF and TRD.

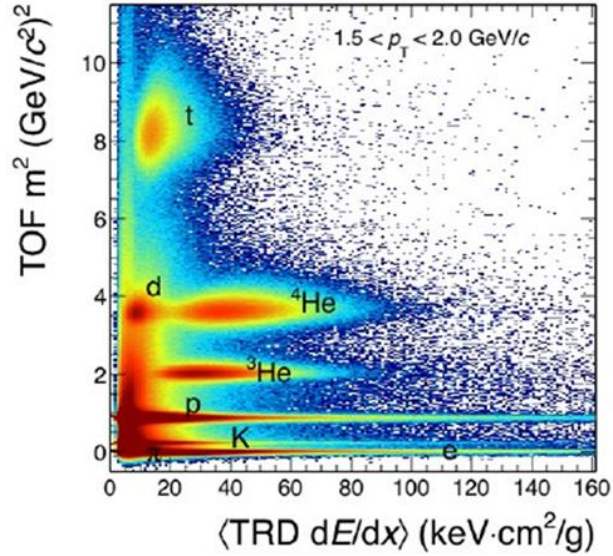

Fig. 26: Squared mass of charged hadrons, derived from their momentum and time-of-flight, as function of the particle energy-loss measured with the TDR.

### 4.6 Collective flow of identified particles

The PSD performance concerning the reconstruction of the reaction plane, and, hence, the measurement of the collective flow of particles, has been simulated for Au-Au collisions with the DCM-QGSM event generator, extended with the Statistical Multi-fragmentation model (SMM) [18]. The GEANT4 code was used to simulate the particle transport through the CBM detector material. The STS and MVD detectors located in the central rapidity region are used for particle's momentum reconstruction and identification, and for centrality estimation. The $\Lambda$ hyperons and the $K_s^0$ mesons have been reconstructed from their decay products with the Kalman Filter (KF) Particle Finder package [15]. The PSD is used for the determination of the reaction plane. The PSD consists of 44 modules of transverse dimensions of $20 \times 20$ cm$^2$. In the detector center, there is a diamond shaped hole with a size of $20 \times 20$ cm$^2$ to let the beam pass through the calorimeter. Figure 27 illustrates the reconstructed direct in-plane flow component $v_1$ of $\Lambda$ hyperons (green dots) and $K_s^0$ mesons (pink dots) using the information from PSD for the estimation of the reaction plane in Au+Au collisions at 12A GeV/c [4]. The dashed areas depict the corresponding $v_1$ reconstructed based on the reaction plane from Monte-Carlo information.

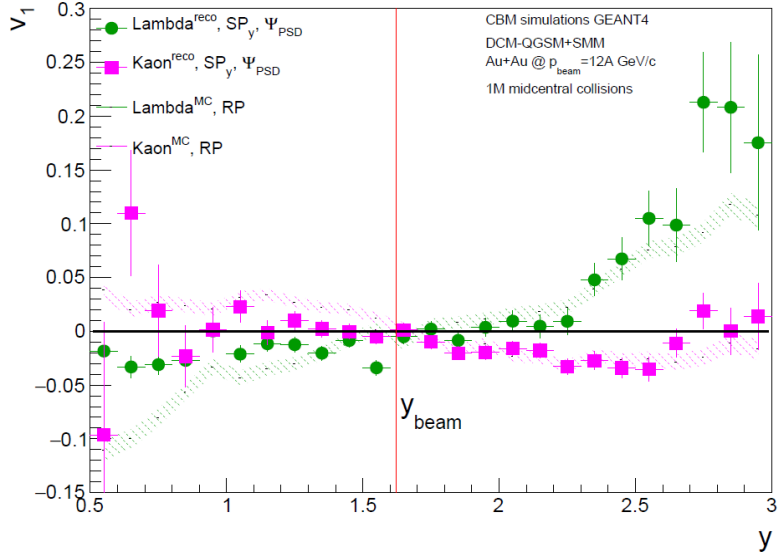

Fig.27: In-plane flow $v_1$ of $\Lambda$ hyperons (green dots) and $K_s^0$ mesons (pink dots) versus rapidity, simulated for Au+Au collisions at 12A GeV/c, and reconstructed based on PSD information. The dashed areas illustrate the corresponding $v_1$ distributions using Monte-Carlo information.

5. Summary

The Compressed Baryonic Matter (CBM) experiment at the future FAIR accelerator centre is designed to investigate the properties of nuclear matter at neutron core densities in the laboratory. The research program includes the exploration of the high-density nuclear matter equation-of-state, the search for new phases of compressed QCD matter, and the study of the strange dimension of the nuclear chart. In order to measure the relevant and partly very rare observables with unprecedented precision and in multi-dimensions, the experiment has to run at extremely high reaction rates, i.e. Au+Au collisions up to 10 MHz. This requires fast and radiation hard detector systems and read-out electronics, followed by a high-speed data acquisition and analysis chain. The CBM experiment comprises a Micro-Vertex Detector and a Silicon Tracking System located in the magnetic field of a superconducting dipole for track reconstruction. The time-of-flight of charged particles will be determined by a large-area detector wall consisting of Multi-Gap Resistive Plate Chambers. Electrons will for be identified by a Ring-Imaging Cherenkov detector together with a Transition Radiation Detector. The RICH detector can be replaced by a muon detection system comprising up to 5 hadron absorbers, two triplets Gas-Electron Multipliers and two triplets Resistive Plate Chambers for particle tracking. The reaction plane will be determined by the Projectile Spectator Detector, a hadronic calorimeter located up to 15 m downstream the target. Most of the observables, such as multi-strange hyperons, do not provide a hardware trigger signal in the high track density environment of a heavy-ion collision. In order to run at very high collision rates, a novel data readout and acquisition system has been developed. The detectors are readout without event correlation by free-streaming ASICs, which mark each detector hit with a time stamp. These hits, carrying information on three space coordinates plus the time coordinate, are used to reconstruct online the particle tracks. After track reconstruction, all charged particles including instable hadrons, electrons and muons are identified, then the particles are assigned to their corresponding events, and finally the events, which contain the relevant information, are selected and stored. These tasks have to be performed online by high-speed algorithms running on up to 40000 cores of a high-performance computing centre.

The CBM experiment at FAIR combines innovative detector and electronics technologies, high-speed algorithms running on modern computer architectures, and a high-intensity heavy-ion beam. These are the necessary prerequisites for a unique research program with substantial discovery potential regarding the fundamental properties of QCD matter.

## Acknowledgment

The development of the CBM experiment is performed by the CBM Collaboration, which consists of more than 470 persons from 58 institutions and 12 countries. The CBM project is supported by the German Ministry of Education and Research, the Helmholtz Association, the Department of Science and Technology and the Department of Atomic Energy, Govt. of India, by the Czech government under MŠMT (MEYS) LM2015049 and LM2018112, by ROSATOM, by the science ministries of Poland, China, Romania, and other national funds of the CBM member institutions. The project receives funding from the Europeans Union's Horizon 2020 research and innovation programme under grant agreement No. 871072. The author acknowledges support from the Ministry of Science and Higher Education of the Russian Federation, grant N 3.3380.2017/4.6 and by the National Research Nuclear University MEPhI in the framework of the Russian Academic Excellence Project (contract No. 02.a03.21.0005, 27.08.2013).